\documentclass[aps,floatfix,twocolumn,groupedaddress,nobalancelastpage]{revtex4}
\usepackage{graphicx}
\usepackage{dcolumn}
\usepackage{bm}
\usepackage{hyperref}
\usepackage{subfigure}
\usepackage[sort&compress]{natbib}
\begin{document}

\title{Cosmic (super)string constraints from 21 cm radiation}
\author{Rishi Khatri}
\email{rkhatri2@uiuc.edu}
\affiliation{Department of Astronomy, University of Illinois at Urbana-Champaign, 1002 W.~Green Street, Urbana, IL 61801}
\author{Benjamin D. Wandelt}
\email{bwandelt@uiuc.edu}
\affiliation{Departments of Physics and Astronomy, University of Illinois at Urbana-Champaign, 1002 W.~Green Street, Urbana, IL 61801}

\date{\today}
\begin{abstract}
We calculate the contribution of
cosmic strings arising from a phase transition in the early universe, or
cosmic superstrings arising from brane inflation, to the cosmic 21 cm
power spectrum at redshifts $z\geq 30$. Future experiments can exploit
this effect to constrain the cosmic string tension $G\mu$ and probe virtually
the entire brane inflation model space allowed by current observations.
Although current experiments with a collecting area of
$\sim 1$ $\rm{km}^2$ will not provide any useful constraints, future
experiments with a collecting area of $10^4-10^6$ $\rm{km}^2$ covering the
cleanest $10\%$ of the sky can in principle constrain cosmic
strings with tension $G\mu \gtrsim 10^{-10}-10^{-12}$ (superstring/phase
transition mass scale $>10^{13}$ GeV).
\end{abstract}
\maketitle

\noindent \emph{Introduction. ---}  
There has been a revival of interest in cosmic
 strings, the line like topological defects of cosmic length,  after it was
 found that they 
 can arise in the superstring theories
 in braneworld inflation scenarios \cite{kibble04}. They are
 also found to be inevitable in a wide class of supersymmetric Grand
 Unified Theories (GUTs) \cite{jean}. Historically 
 cosmic strings  were found to form in GUTs during
 phase 
 transitions in the early Universe along with  the other topological defects
 like monopoles and domain walls \cite{kib76}. Unlike monopoles and
 domain walls, which 
  very quickly dominate the energy density of the
 Universe if formed after inflation, strings approach a scaling solution and can remain
 sub-dominant \cite{kib85}. In brane inflation only cosmic strings and no
 monopoles or domain walls are
 produced \cite{sar02}. 

Cosmic strings were proposed as a mechanism
 for generating
 the primordial fluctuations which later grew to
 form the large scale structures we see today \cite{zel80}.  The Cosmic
 Microwave 
 Background (CMB) anisotropies and the matter power spectrum
 arising from the fluctuations seeded by cosmic strings are very different
 from those generated from inflation \cite{guth81}. Inflation
 just prescribes the initial fluctuations at the end of inflation generated
 once and for all, which then 
 just evolve.
 Cosmic strings generate fluctuations throughout the history
 of the Universe. One effect of this is that the fluctuations generated at
 different times add up out of phase and wash out the acoustic oscillations
 in the CMB. The discovery of the acoustic peaks by CMB experiments
 \cite{bern00,wmap3} was a major 
 success for inflation and ruled out cosmic strings as the dominant
 mechanism for seeding the cosmic perturbations.   A sub-dominant
 contribution from cosmic strings to the cosmic perturbations is still not
 ruled out  with the current constraint
 being $G\mu \lesssim 10^{-7}$ for classical strings \cite{pog05}. Cosmic strings if
 discovered either through their 
 gravitational lensing effects \cite{sazhin07}, through the gravitational
 waves produced at string cusps or decaying loops \cite{damour} or through their effect
 on the CMB and the
 matter power spectrums will provide insight into the fundamental physics
 at high energies
 which is beyond the reach of currently planned terrestrial
 experiments. 

21 cm radiation from $z\geq 30$ is an excellent probe of the state of the
Universe at that time. This radiation can probe much smaller scales than
the CMB, in the redshift range $30\leq z\leq 200$, and provides a three dimensional view
of the Universe before reionization \cite{loeb}. It has been shown to be an
excellent probe of the fundamental physics like variation of constants
(fractional variation in the fine structure constant of $\lesssim 10^{-5}$ with
$10^4\rm{km}^2$ collecting area),
non-Gaussianity from inflation (non-Gaussianity parameter $f_{nl} \sim 0.01-1$), dark matter and inflationary fundamental
physics 
\cite{cor,kw,vald07}. In this \emph{Letter} we show that it
can, in principle, put 
unprecedented tight constraints on the cosmic string contribution to the
perturbations in the matter or equivalently on the string tension $G\mu$,
and possibly other parameters, 
which translates into constraints on the GUTs and the superstring theory.
$G$ is the gravitational 
constant and $\mu$ is the string mass per unit length so that $G\mu$ is
dimensionless.

\noindent \emph{Cosmic strings. ---}
 Cosmic strings arise in GUTs and superstring theories whenever there is a phase transition in the Universe
 if the vacuum manifold contains unshrinkable loops, e.g. U(1).  The superstring theory
 can produce a variety of cosmic strings, which can be fundamental
 (F-)strings or D-strings produced during annihilation of D-branes. The
 string tension for these strings in brane inflation models is $10^{-12}\lesssim G\mu \lesssim
 10^{-6}$ \cite{sar02,jones02,pol}.
 Just
 like the classical cosmic strings from GUTs, they are wiggly, can
 intercommute and form 
 loops which can decay into gravitational radiation or elementary
 particles. The
 main difference from the classical cosmic strings of GUTs is that their
 intercommuting probability can be less than unity. Also different kinds of
 strings can form bound states \cite{tye05}. The superstring
 theory string networks have scaling solutions \cite{sakel} just like
 the classical strings \cite{kib85} i.e. the total length of the
 strings inside a horizon volume is proportional to the horizon size at any
 time and the string energy density
 is a constant fraction of the dominant energy density component of the
 Universe in the radiation dominated as well as matter dominated eras,
 preventing them from dominating.
 This also makes possible to construct simpler models of string networks which can than be
 used to study their impact on cosmology. Also since the perturbations
 produced by cosmic strings are independent of the inflationary initial
 conditions, the two kind of perturbations can be evolved independently and
 the resulting power spectra for CMB or matter added together to get the
 total power spectrum.

The long wiggly cosmic strings have
  a structure that resembles a random walk on scales
 larger than the horizon but are straight on small scales
 \cite{shellard}. We use the CMBACT code developed by Pogosian and Vachaspati \cite{pog99}
 which is based on CMBFAST \cite{cmbfast}.  Wiggly strings are modeled in the code as independent pieces of
 small strings whose length is taken to be of the order of the correlation
 length of the pieces of strings derived from numerical simulations.
The intercommuting probability ($P$) is taken to be unity. The wiggliness on small scales results in the
effective string tension and mass per unit length of the string observed by
a distant observer to differ \cite{carter89} with the equation of
state $\tilde{U}\tilde{T}=\mu^2$,  where $\tilde{U}$ and $\tilde{T}$ are
the effective mass per unit length and tension of the wiggly string. Thus at scales greater than the
scale of wiggles, the matter around experiences a Newtonian gravitational potential
in addition to the deflection due to the conical space around the string
\cite{vachas91}. The wiggliness can be controlled in CMBACT by a wiggliness
factor $\alpha$ defined by the
 equations $\tilde{U} = \alpha \mu$ and $\tilde{T}=\mu/\alpha$ \cite{pog99}.

\noindent \emph{21 cm radiation--} After recombination most of the baryonic matter is
in the 
form of neutral atomic hydrogen and helium with a very small residual
ionization. This small fraction of electrons couples the CMB to the matter
up to $z\sim 500$ through Compton scattering and maintains
the gas at the same temperature as the CMB. Around $z\sim 500$ the Compton
scattering timescale becomes larger than the Hubble time and the gas
decouples from the CMB and cools adiabatically thereafter. The ground state
of
hydrogen atom has a hyperfine splitting with an energy difference of $T_{\star}=0.068$
K. This corresponds to the $\nu_{21}\sim 1420$ MHz rest frequency or
$\lambda_{21}\sim 21$ cm rest 
wavelength. The hydrogen gas will absorb or emit photons of this energy
depending on the population levels in the two states which is best
parameterized by the spin temperature $T_s$ defined by the equation
$n_t/n_s = g_t/g_se^{-T_{\star}/T_s}$. Here $n_t$ and $n_s$ are the
number densities of hydrogen atoms in the excited triplet state and the ground
singlet state respectively, $g_t=3$ and $g_s=1$ are the corresponding
statistical weights. The population levels during
$z\geq 30$ can change through collisions or through emission and absorption of
CMB photons. Initially the collisions dominate over the radiative process and
$T_s$ follows the gas temperature $T_g$. At late times the density of gas
becomes too low for collisions to be effective and $T_s$ approaches the CMB
temperature, $T_{\gamma}=2.725(1+z)$. Thus in the redshift range of about
$500\geq z\geq 30$, $T_s<T_{\gamma}$ and we have a net absorption of the CMB
\cite{loeb,bhardwaj}. At $z\sim 30$ the first stars are born and the evolution of the
Universe enters non-linear regime. We will focus on the redshift range $200\geq z\geq 30$ in this
\emph{Letter}, which corresponds to the observed frequency of $\sim 7.1\rm{MHz}\leq \nu \leq 
47.3\rm{MHz}$.  

The observed intensity $I_{\nu}$ can be
expressed in terms of the brightness temperature using the Rayleigh-Jeans
formula, $T_b=I_{\nu}c^2/2k_B\nu^2$, with $T_b$ given by
\cite{field} 
\begin{eqnarray}
\label{tb}T_b = \frac{\left(T_s - T_{\gamma}\right)\tau}{\left(1+z\right)},
\hspace{10 pt}
\tau=\frac{3c^3\hbar A_{10}n_H}{16k_B\nu_{21}^2(H + (1+z)\frac{dv}{dr}) T_s},\nonumber
\end{eqnarray}
where $n_H$ is the number density of
atomic Hydrogen, $A_{10}$ is the Einstein A
coefficient for spontaneous emission, $c$ is the speed of light in vacuum,
$k_B$ is Boltzmann's constant and $\hbar=h/2\pi$ with $h$ being
Planck's constant. $H$ is the Hubble parameter, $r$ is
the comoving distance to redshift $z$ and $v$ is the
peculiar velocity along the line of sight. We ignore the contribution of
the vector modes to peculiar velocities since it is sub-dominant ($< 1 \% $) compared to
the scalar mode contribution at scales of interest ($l>1000$). All quantities except the
physical and atomic constants are functions of $z$.
There will be spatial fluctuations in $T_b$ and $T_s$ caused by
the fluctuations in $n_H$ and $T_g$, which in the redshift range of interest are
related  to the linearly evolved primordial perturbations in standard
inflationary cosmology. We will see below that the cosmic strings, if they
exist, can add a significant contribution to these fluctuations. 

Expanding the fluctuations in the brightness temperature $\delta
T_b=T_b-\bar{T_b}$, where $\bar{T_b}$ is the mean brightness temperature,
in spherical harmonics, we get the angular power spectrum $C_l(z)=\langle
a_{lm}a_{lm}^*\rangle$, where $a_{lm}$ are the coefficients in the
spherical harmonic expansion of $\delta T_b$. Following \cite{bhardwaj} we
can write, 
\begin{eqnarray}
C_l^i(z)=\int \frac{d^3k}{\left(2\pi\right)^3}P^i(k,z)S_l(k,z),\nonumber
\end{eqnarray}
where $P(k,z)$ is the baryon (Fourier) power spectrum, index $i=ad$ or $cs$
for adiabatic perturbations from inflation or perturbations from cosmic
strings respectively. We have incorporated the 21 cm physics into  $S_l(k,z)$
 \cite{bhardwaj}. 
\begin{figure}
\includegraphics{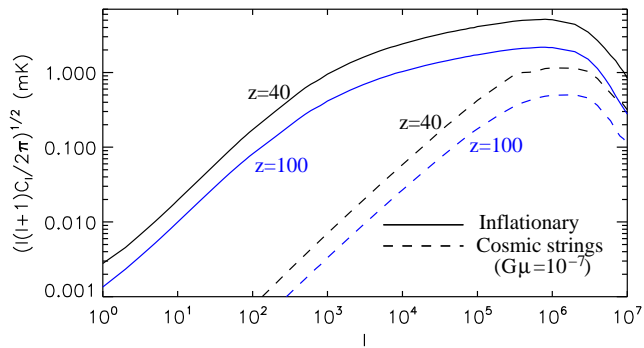}
\caption{\label{cl}Angular power spectra from inflationary adiabatic
 initial conditions (COBE normalized) and cosmic strings.} 
\end{figure}

\noindent \emph{Results--} We calculate $P^{ad}(k,z)$ using CMBFAST \cite{cmbfast}
and $P^{cs}(k,z)$ using CMBACT \cite{pog99}. The cosmological parameters
are from WMAP3 assuming $\Lambda CDM$ cosmology \cite{wmap3}. For the cosmic string model in
CMBACT we use initial rms velocity of $0.65$, wiggliness factor in the
radiation era of $1.9$ and initial correlation length of $0.13$ times the
initial conformal time, motivated by numerical simulations  \cite{martins2002,shellard}. The intercommuting probability of strings is taken to be unity, which
means classical strings. The effect of smaller intercommuting probability
will translate into a denser network which will make $G\mu$ smaller for the same
amplitude of string generated perturbations.  Our predicted constraints
on $G\mu$ are therefore  conservative.

The angular power spectra for
two redshifts are plotted in Fig. \ref{cl} for $G\mu =
10^{-7}$ along with the inflationary spectra. One important feature of the cosmic string power spectrum is that
it turns over at smaller scales compared to the inflationary adiabatic
power spectrum. This is due to the fact that the strings
continue to generate perturbations actively at all times.
\begin{figure}
\includegraphics{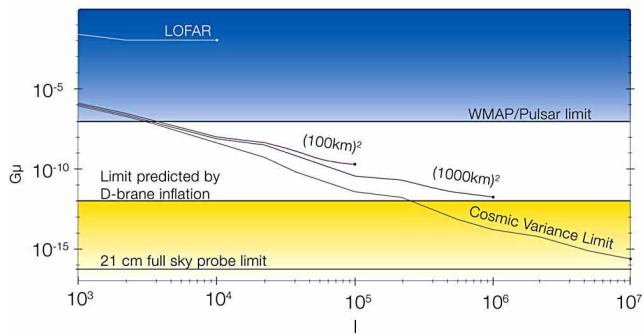}
\caption{\label{fish} Constraints from current and future experiments on $G\mu$ for a sky fraction of
 $10\%$, bandwidth of 0.4 MHz and integration time of 3 years. Also marked is the constraint on $G\mu$ achievable by a full sky cosmic
 variance limited experiment for the parameters assumed in the present calculation.}
\end{figure}
To estimate the constraints possible on $G\mu$ from these observations we
calculate the Fisher matrix $F_{\theta}$ for the parameter
$\theta=(G\mu/10^{-7})^2$, assuming that all the other parameters are
known.
\begin{eqnarray}
F_{\theta}^i=\sum_{z}\sum_{l=1}^i{\frac{f_{sky}(2l+1)}{2}\left(C_l^{ad}(z)+
C_l^N(z)\right)^{-2}\left(\frac{\partial C_l(z)}{\partial \theta}\right)^2}, \nonumber
\end{eqnarray}
where $C_l=C_l^{ad} + C_l^{cs}$, $f_{sky}$ is the sky fraction observed and the sum is over all redshift slices and
$l$ up to a maximum $l=i$. We use the fact that the bandwidth of any experiment is finite  only to
estimate the noise $C_l^N$ and the number of redshift slices available and
ignore  its  damping effect on $C_l$ \cite{lewis}. Bandwidth of $0.4$ MHz
ensures that the slices are seperated by $\gtrsim  r/l$ for $l\gtrsim 1000$ and are thus
uncorrelated.

The noise in
the frequency range $7<\nu<47$ MHz is dominated by the sky temperature
which follows a power law, $(C_l^N)^{1/2}\propto T_{sky} \propto
\nu^{-2.5}$ \cite{22}. For LOFAR \cite{lofar}
 we use the noise estimates of \cite{zald} at z=10, scaled to
low frequencies using the above mentioned power law with a bandwidth
$\Delta \nu = 0.4$MHz and integration time of 3 years. These results also apply
to LWA \cite{lwa} which has similar specifications as LOFAR.
Fig. \ref{fish} shows  the constraints on $G\mu$ assuming an error of $1/2(F_{\theta}^l)^{1/2}$ (the factor of 2 comes
in when we convert the error on $\theta$ to the error on $\theta^{1/2}$) for
LOFAR, two futuristic experiments and
 the cosmic variance limit assuming that the foregrounds can be
removed at required precision \cite{santos,zald}. Most of the
information on $G\mu$ is at $l>10^4$ as is clear from an
inspection of the power spectra also. 
For the cosmic variance limit we use the fact that there are more
independent modes at
high l, $\Delta \nu \approx \nu^2rH/\nu_{21}lc$ \cite{loeb} to
calculate the number of redshift slices.
The curve 
 labeled $(100\hspace{2 pt}\rm{km})^2$ corresponds to a futuristic
 telescope of size $100$ km and  collecting area of $10^4$ $\rm{km}^2$ that
 will reach out to $l\sim 10^5$ and $G\mu \sim 10^{-10}$. A $(1000\hspace{2
   pt}\rm{km})^2$ telescope will be needed to constrain $G\mu
 \sim 10^{-12}$.  Such a telescope might be possible not only on Earth but also
  in space \cite{space} or on
the far side of Moon \cite{carilli}. 
 To reach $G\mu \sim 10^{-14}$ will
require a collecting area of $10^{12}\hspace{2 pt}\rm{km}^2$. Reaching the
cosmic variance limit of $\sim 10^{-15}$ may not be possible
because the scattering of radio waves by the ionized interstellar medium
will limit the smallest angular scales ($\propto \nu^{-2}$) that can be observed
\cite{cohen}. In particular the $l >
10^6$ modes will not be available at all redshifts.

It is clear from Fig. \ref{fish} that the information content of the 21 cm
signal is huge and can in principle constrain $G\mu \sim 10^{-16}$, if we
have a cosmic variance limited measurement on the full sky. This
corresponds to a phase transition energy scale of $10^{11}$ GeV for GUT
theories. If we take $\mu \sim 2M_s^2$ for D-brane strings, this means
bounds on the superstring mass scale, $M_s$, down to $\sim 10^{11}$ GeV
\cite{sar02}. In reality only a fraction of the sky would be
available due to our being confined to the galaxy and small scale modes
may not be available because of the interstellar scattering of radio waves, but even then the 21 cm
signal has impressive constraining power over $G\mu$.  The power at small scales
due to cosmic strings is in fact underestimated in this linear
calculation. Cosmic strings generate wakes behind them, which have a
density contrast of unity, that is not taken into account here and
which would enhance the power due to cosmic strings at small scales through
non-linear gravitational evolution.
This is just the information contained in the power spectrum. Higher
order correlations will 
provide additional discriminating power to check for the signatures of
perturbations seeded by a network of cosmic strings. The non-Gaussianities
due to cosmic
strings would be larger due to the highly non-linear nature of the
perturbations at small scales  and different from those produced during
inflation \cite{cor}. 
This impressive constraint is due the fact that there are large number of modes available at high l
so that the statistical errors become very small. 

High redshift 21 cm observations thus provide a rare observational window into
the superstring theory and supersymmetric grand unified theories through cosmic strings.

\noindent \emph{Acknowledgments. ---} We thank Xavier Siemens and Richard
Battye for discussions on the subject. We also thank Levon Pogosian and
Tanmay Vachaspati for making their
code publicly available and answering our questions related to it. We thank the referee for comments that improved the
presentation of our results and Nikita Sorokin for assistance in preparing Figure 2. We thank
Max Planck Institute for Astrophysics for their hospitality where most of this work was done. Ben Wandelt was
supported by the Alexander von Humboldt foundation's  Friedrich Wilhelm Bessel research award.
\bibliographystyle{apsrev}
\bibliography{kw_08}
\end{document}